\documentclass[journal=jpcbfk,manuscript=article]{achemso}

\usepackage[T1]{fontenc}
\usepackage{chemfig}
\usepackage{float}
\usepackage[ruled,vlined]{algorithm2e}
\usepackage{listings}
\usepackage{algorithmic}

\author{Mikita M. Misiura}
\affiliation{Department of Chemistry and Center for Theoretical Biological Physics, Rice University, Houston, TX 77005, USA}
\author{Alexander M. Berezhkovskii}
\affiliation{Mathematical and Statistical Computing Laboratory, Office of Intramural Research, Center for Information Technology, National Institutes of Health, Bethesda, MD 20892}
\author{Sergey M. Bezrukov}
\affiliation{Section on Molecular Transport, Eunice Kennedy Shriver National Institute of Child Health and Human Development, National Institutes of Health, Bethesda, MD 20892}
\author{Anatoly  B.  Kolomeisky}
\affiliation{Department of Chemistry and Center for Theoretical Biological Physics, Rice University, Houston, TX 77005, USA}
\alsoaffiliation[Rice University]{Department of Chemical and Biomolecular Engineering, Rice University, Houston, TX 77005, USA}
\alsoaffiliation[Rice University]{Department of Physics and Astronomy, Rice University, Houston, TX 77005, USA}

\email{tolya@rice.edu}

\title{Surface-Facilitated Trapping by Active Sites: From Catalysts to Viruses}

\begin{document}

\begin{abstract}

Trapping by active sites on surfaces plays important roles in various chemical and biological processes, including catalysis, enzymatic reactions, and viral entry into host cells. However, the mechanisms of these processes remain not well understood, mostly because  the  existing theoretical descriptions are not fully accounting for the role of the surfaces. Here we present a theoretical investigation on the dynamics of surface-assisted trapping by specific active sites. In our model, a diffusing particle can be occasionally reversibly bound to the surface and diffuse on it before reaching the final target site. An approximate theoretical framework is developed, and its predictions are tested by Brownian Dynamics computer simulations. It is found that the surface diffusion can be crucial in mediating the association to active sites. Our theoretical predictions work reasonably well as long as the size of the active site is much smaller than the overall surface area. Potential applications of our method are discussed.

\end{abstract}

 
\newpage

\section*{Introduction}

There are multiple chemical and biological processes that involve association of particles to specific reactive sites that are located on surfaces. Examples include catalytic reactions on heterogeneous catalysts\cite{somorjai2010introduction} and  enzymatic processes on active sites located on the protein surfaces.\cite{copeland2000enzymes} Analysis of these systems typically considers the molecular surfaces  as inactive and not participating in these processes. However, the surfaces of solid-state catalysts and enzymatically active proteins  are clearly not inert, and the reacting species interact with them affecting the overall outcome of the process.  These interactions might involve hydrogen bonds, electrostatic forces and hydrophobic interactions, and they might achieve significant strengths that could strongly modify the overall dynamics of the process. For instance, recent all-atom molecular dynamics simulations of association of small ligands with dihydrofolate reductase proteins explicitly demonstrated the effect of surface facilitation on the kinetics.\cite{nerukh2012ligand}

A relatively new example of the reaction facilitated by surface diffusion is channel-catalyzed transport of membrane-bound peripheral proteins.\cite{rostovtseva2015alpha,hoogerheide2018real}  It was recently shown that $\alpha$-synuclein – the cytosolic protein implicated in Parkinson disease – enters mitochondrial periplasmic space through a process involving reversible binding to the outer membrane of mitochondria as the first crucial step.  Protein then diffuses on the membrane surface and either escapes into the bulk or is captured by the $\beta$-barrel nanopore of the voltage-dependent anion channel, VDAC.  The capture of a membrane-bound $\alpha$-synuclein molecule happens via its highly negatively charged C-terminal tail entering the nanopore with a net-positive charge.  The on-rate of the process is defined by the surface concentration of $\alpha$-synuclein molecules and their conformation, both of which are affected by the membrane lipid composition.\cite{jacobs2019probing,hoogerheide2020tunable}  Since the lipid effects are strong it was suggested that changes in the lipid composition of the host membrane could constitute a potent regulatory mechanism of the $\alpha$-synuclein interaction with VDAC, including its VDAC-catalyzed translocation into the periplasmic space where $\alpha$-synuclein obstructs the normal functioning of the machinery of the inner membrane integral proteins of mitochondria.

The crucial involvement of the membrane surface in the $\alpha$-synuclein capture was clearly demonstrated in experiments with another $\beta$-barrel nanopore,\cite{gurnev2014alpha} formed by $\alpha$-hemolysin which is known to have a pronounced structural asymmetry.\cite{song1996structure}  On one side of the membrane, it protrudes above the membrane surface for about 5 nanometers (the cap side of the channel structure), while on the other side it is flush with the surface (the stem side).  It was found that the protruding, cap-side opening of the channel displays several orders of magnitude lower capture rates, presumably capturing $\alpha$-synuclein only from the bulk solution, compared to the rates of capture from the membrane surface by the other, stem-side opening.  This huge difference in the capture on-rates at otherwise symmetric conditions, as well as the highly expressed sensitivity of the on-rate to the membrane lipid composition, support the picture according to which the $\alpha$-synuclein molecule first binds to the membrane and then diffuses over its surface before being captured by the nanopore or released back to the bulk solution.

Another example of the complex process where interactions with the surfaces are important is the entry of viruses into host cells.\cite{sobhy2017comparative} Recent experimental studies indicated that the viral penetration is a very complex process that involves multiple pathways. It was shown that viruses often bind first to a random spot on the cell's surface and then diffuse along the surface before binding to the appropriate receptor that allow them to enter into the cell. \cite{rothenberg2011single, marsh2006virus} High-resolution fluorescence microscopy and single particle tracking techniques have illustrated how $\lambda$ phage is able to first bind to a random location on the \textit{E. coli} surface and then use complicated target-searching process to reach one of the cell's poles.\cite{rothenberg2011single} It was also shown that sometimes multiple receptors are needed for a virus to enter the cell, which might lead to even longer and more complex  diffusion processes.\cite{gibbons2010diffusion}

The problem of association of small ligands with specific active sites on the surfaces has been investigated before using various theoretical tools.\cite{berg1985orientation,zhou2004enhancement,zhou2010rate,shin2018surface} A diffusion limited association, when the reaction of binding to the surface (both specific and non-specific) is instantaneous, has been considered by Berg.\cite{berg1985orientation} In the elegant approach by Zhou and Szabo, the effect of the surface was modelled as a short-range attraction potential, leading to analytical results for different systems.\cite{zhou2004enhancement,zhou2010rate}  The surface-assisted association has been analyzed as a dynamic search process by one of us,\cite{shin2018surface} but the analytical results have been obtained only in certain limiting cases. It should be noted here that the enhancement of the association rates due to the presence of the surface is an example of more general dynamic target search facilitation which is accomplished by switching the participating particles between different dynamic regimes with different dimensionalities. In this case, particles are alternating between 3D bulk motion and 2D surface sliding. It is worth mentioning that significant progress has been achieved  earlier in understanding the mechanisms of facilitated diffusion,\cite{mirny2009protein,kolomeisky2011physics,sheinman2012classes,koslover2011theoretical,bauer2013vivo,shvets2018mechanisms,kochugaeva2016conformational,felipe2021dna,cencini2018energetic}  where the dynamics switches between 1D and 3D regimes.  This 3D/1D facilitation mechanism has been extensively discussed due its importance for protein target search on DNA that initiates all major biological processes. These ideas were initially stimulated by pioneering work of Adam and Delbruck on dimensionality reduction and diffusion in biological systems.\cite{adam1968structural}

In this work, we propose a new theoretical framework to describe the surface-assisted molecular/particle associations to specific sites. It can be viewed as a generalization of the Smoluchovskii-Collins-Kimball method that accounts for nonspecific interactions in the system. We model one of the binding partners as an immobile partially absorbing sphere with a small perfectly absorbing patch that corresponds to the active site, and the second binding partner is viewed as a diffusing point particle that might bind to the surface. It can diffuse then along the surface until it finds the active site, or it can dissociate back into the bulk solution. If the size of the active site is  small, the  overall binding process is considered as trapping by the  uniform partially absorbing sphere. This allows us to derive explicit formulas for dynamic properties of the process. Extensive Brownian Dynamics (BD) simulations are utilized then to test our theoretical predictions and the range of parameters where our approximate method works.

\section*{Theoretical Method}

To obtain a theoretical description of the surface-assisted trapping of particles to active sites, let us consider a system schematically shown in Fig. 1. We have an immobile sphere of radius $R$ surrounded by the medium with diffusing point particles. When the sphere is perfectly absorbing, a particle touching its surface is instantly trapped. In this case, the steady-state flux of trapped particles is given by the Smoluchovski formula,\cite{smoluchowski1916drei}
\begin{equation}\label{J_Sm}
    J_{Sm}=k_{Sm} c_{\infty}, \quad k_{Sm}=4 \pi D R,
\end{equation}
where $k_{Sm}$ is the Smoluchowski rate constant, $D$ is the particle diffusivity and $c_{\infty}$ is the concentration of particles far away from the sphere. 

If the sphere is only partially absorbing, the steady-state flux of trapped particles is lower than $J_{Sm}$ since a particle touching the sphere has a chance to escape  back to the bulk solution. This flux is given by the Collins-Kimball theory,\cite{collins1949diffusion}
\begin{equation}\label{J_CK}
    J_{CK}=k_{CK} c_{\infty}= J_{Sm} P_{tr}^{CK},
\end{equation}
where $k_{CK}$ is the Collins-Kimball rate constant, and $P_{tr}^{CK}$ is the trapping probability for a particle that starts from the surface of the sphere. The Collins-Kimball rate constant is given by
\begin{equation}\label{k_CK}
    k_{CK}=\frac{k_{Sm}k_{tr}}{k_{Sm}+k_{tr}}=k_{Sm} P_{tr}^{CK},
\end{equation}
where $k_{tr}=4\pi R^{2} \kappa_{tr}$ is the rate constant in the trapping-controlled regime (when $D \rightarrow \infty$), with $\kappa_{tr}$ denoting the surface trapping rate ($\kappa_{tr}=0$ and $\infty$ for perfectly reflecting and absorbing surfaces, respectively). As follows from Eqs. (\ref{J_Sm}), (\ref{J_CK}) and (\ref{k_CK}), the trapping probability, defined as $P_{tr}^{CK}=J_{CK}/J_{Sm}=k_{CK}/k_{Sm}$, is equal to
\begin{equation}\label{P_CK}
    P_{tr}^{CK}=\frac{k_{tr}}{k_{Sm}+k_{tr}}=\frac{1}{1+D/(R \kappa_{tr})}.
\end{equation}
As $D/(R \kappa_{tr}) \rightarrow 0$, the trapping probability tends to unity and $k_{CK}$ approaches $k_{Sm}$. In the opposite limit when $D/(R \kappa_{tr})\rightarrow \infty$, the probability $P_{tr}^{CK}$ tends to zero as $R \kappa_{tr}/D$ and $k_{CK}$ approaches $k_{tr}$.

\begin{figure}
    \centering
    \includegraphics[clip,width=1.0\linewidth]{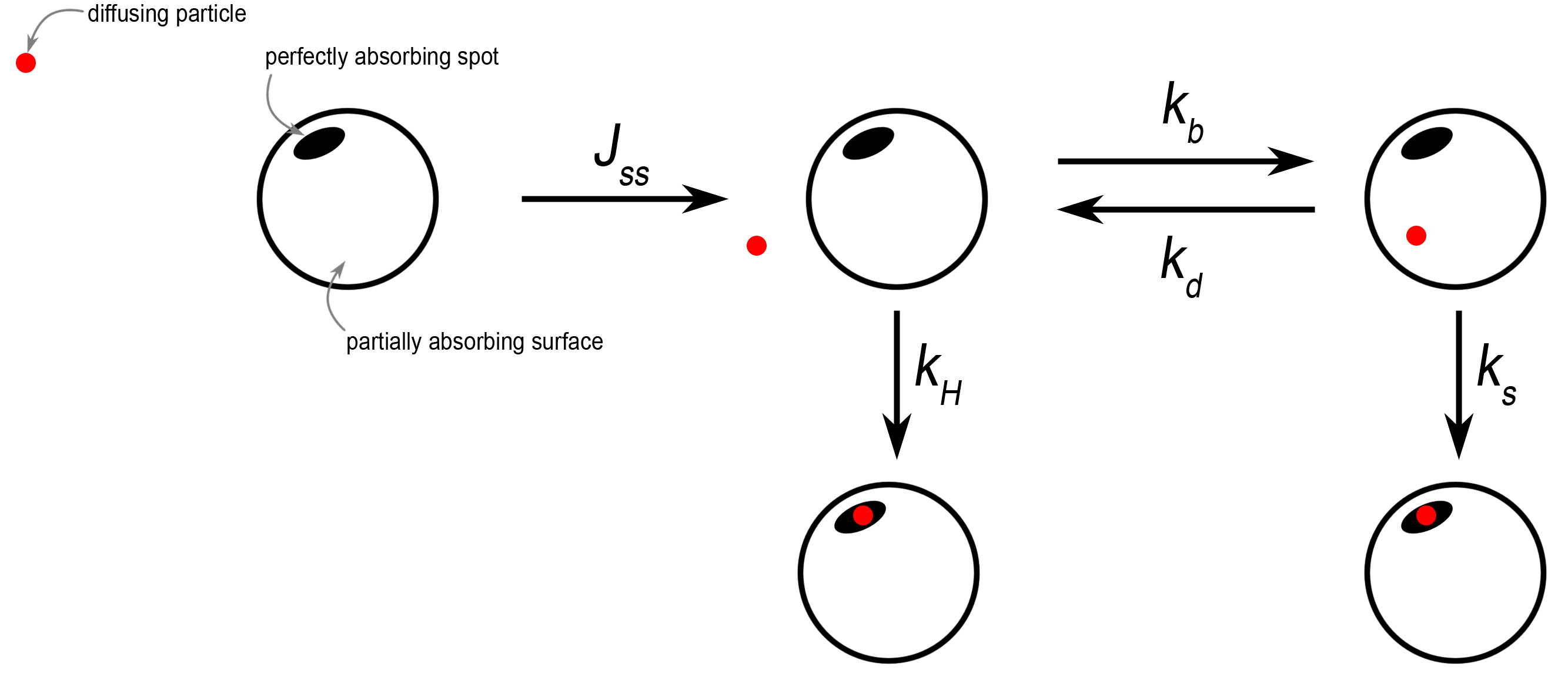}
    \caption{A schematic view of dynamic processes during the particle association to an active site located on a partially absorbing spherical surface. Details are explained in the text.}
    \label{fig:scheme}
\end{figure}

However, in a more realistic situation when the trapping ability is not uniform over the surface of the sphere (Fig. 1), we can introduce an effective trapping probability $P_{tr}$ and an effective rate constant $k$, such that the steady-state flux of trapped particles can be written as
\begin{equation}\label{J_ss}
    J_{ss}=J_{Sm} P_{tr} = k c_{\infty}, \quad k=k_{Sm} P_{tr}.
\end{equation}
By analogy with Eq. (\ref{P_CK}), we can use $P_{tr}$ to introduce an effective uniform trapping rate of the surface, $\kappa_{eff}$, by the following relation,
\begin{equation}
    P_{tr}=\frac{J_{ss}}{J_{Sm}}=\frac{k}{k_{Sm}}=\frac{1}{1+D/(R \kappa_{eff})},
\end{equation}
which leads to 
\begin{equation}\label{kappa_eff}
    \kappa_{eff}=\frac{D P_{tr}}{R(1-P_{tr})}.
\end{equation}
This allows us to map the problem of nonuniform trapping by a spherical surface to an effective Collins-Kimball problem of trapping by the uniform partially absorbing sphere characterized by the effective trapping rate $\kappa_{eff}$. To determine $\kappa_{eff}$, we have to find the steady-state flux, from which one can calculate the effective rate constant and trapping probability using the relations given in Eq. (\ref{J_ss}) and eventually the expression for $\kappa_{eff}$ by Eq. (\ref{kappa_eff}).

Applying the "boundary homogenization" approach outlined above to trapping of diffusing particles by a small absorbing disk of radius $a$ located on the surface of otherwise reflecting sphere of radius $R$ ($a \ll R$: see Fig. 1), we can write the steady-state flux as
\begin{equation}
    J_{ss}=k_{HBP} c_{\infty}=J_{Sm} P_{tr}^{HBP}.
\end{equation}
Here $k_{HBP}$ is the Hill-Berg-Purcell rate constant that describes trapping of diffusing particles by an absorbing disk of radius $a$ located on the otherwise reflecting flat wall,\cite{hill1975effect,berg1977physics}
\begin{equation}\label{kHBP}
    k_{HBP}= 4Da=k_{Sm} P_{tr}^{HBP},
\end{equation}
and $P_{tr}^{HBP}=J_{ss}/J_{Sm}=k_{HBP}/k_{Sm}$ is the probability to be trapped by the disk for a particle whose starting position is uniformly distributed over the surface of the sphere, including the disk, 
\begin{equation}
    P_{tr}^{HBP}=\frac{a}{\pi R}.
\end{equation}
Using this, we can introduce the effective trapping rate $\kappa_{eff}^{HBP}$ which, according to Eq. (\ref{kappa_eff}), is given by
\begin{equation}
    \kappa_{eff}^{HBP}=\frac{D P_{tr}^{HBP}}{R(1-P_{tr}^{HBP})}=\frac{Da}{R(\pi R-a)} \simeq \frac{Da}{\pi R^{2}}
\end{equation}
as $a \ll R$.

Now we generalize the above considerations to a more realistic case where diffusing particles can reversibly bind to the surface of the sphere outside the disk and diffuse on this surface with a diffusivity $D_{s}$ (Fig. 1). This allows particles to reach the disk not only via direct association from the bulk solution but also by the surface diffusion, leading to the overall increase in the steady-state flux and hence the effective rate constant. The steady-state flux of particles trapped by the disk is the sum of the fluxes coming to the disk from the bulk and from the surface, denoted by $J_{ss}^{(b)}$ and $J_{ss}^{(s)}$, respectively. Since the disk is small, $a \ll R$, its presence affects the particle concentration only in close vicinity of the disk. With this in mind,, we assume that the flux $J_{ss}^{(b)}$ can be well approximated as
\begin{equation}
    J_{ss}^{(b)}=k_{HBP} c(R),
\end{equation}
where $c(R)$ is the steady-state concentration of unbound particles near the surface of the sphere.

In addition, we assume that the flux $J_{ss}^{(s)}$ is the product of the mean number $N_{s}$ of bound particles diffusing on the surface at the stationary state and the rate constant $k_{s}$ that describes the trapping of these particles by the disk,
\begin{equation}\label{J_sss}
   J_{ss}^{(s)}=k_{s} N_{s}. 
\end{equation}
Since the disk is small, $k_{s}$ is the inverse mean fist-passage time $\tau_{s}$ of a particle diffusing on the surface to the disk boundary, conditional on that the particle starting point is uniformly distributed over the surface. This mean first-passage time is given by\cite{bloomfield1979diffusion}
\begin{equation}
    \tau_{s}=\frac{R^{2}}{D_{s}} f(a/R), \quad f(z)= 2 \ln(2/z)-1.
\end{equation}
Thus, the rate constant $k_{s}$ is
\begin{equation}
    k_{s}=\frac{1}{\tau_{s}}=\frac{D_{s}}{R^{2} f(a/R)}.
\end{equation}

The steady-state number of particles diffusing on the surface is proportional to the stationary concentration $c(R)$ of unbound particles near the surface. Let $k_{b}$ and $k_{d}$ be the rate constants for the particle binding to and dissociation from the surface, respectively, as illustrated in Fig. 1. We use these rate constants and the trapping rate constant $k_{s}$ to write the balance equation for the steady-state number of particles diffusing on the surface,
\begin{equation}
    (k_{d}+k_{s})N_{s}=k_{b} c(R).
\end{equation}
Thus, we have
\begin{equation}
    N_{s}=\frac{k_{b}}{k_{d}+k_{s}} c(R).
\end{equation}
This allows us to write the flux $J_{ss}^{(s)}$, Eq. (\ref{J_sss}), in terms of the concentration $c(R)$,
\begin{equation}
    J_{ss}^{(s)}=k_{s} N_{s}=\frac{k_{s}k_{b}}{k_{d}+k_{s}} c(R).
\end{equation}
Summing up the two contributions to the steady-state flux of the particles trapped by the disk, we obtain
\begin{equation}\label{J_ss1}
    J_{ss}=J_{ss}^{(b)}+J_{ss}^{(s)}=\left( k_{HBP}+\frac{k_{s}k_{b}}{k_{d}+k_{s}} \right) c(R).
\end{equation}

This steady-state flux is maintained by the bulk  diffusion of particles to the sphere from infinity. Assuming that the angular anisotropy of the particle concentration in the bulk is small and, therefore, can be neglected, and denoting this concentration at distance $r$ from the center of the sphere ($r>R$) by $c(r)$, one can write the steady-state flux in the bulk as
\begin{equation}
    J_{ss}=4 \pi r^{2} \frac{d c(r)}{dr}, \quad r>R.
\end{equation}
Solving this equation subject to the boundary condition $c(r)|_{r \rightarrow \infty}=c_{\infty}$, we find that
\begin{equation}
    c(r)=c_{\infty}-\frac{J_{ss}}{4 \pi D r}.
\end{equation}
This expression is now used to find the concentration $c(R)$ of unbound particles near the surface of the sphere:
\begin{equation}
    c(R)=c_{\infty}-\frac{J_{ss}}{4 \pi D R}=c_{\infty}-\frac{J_{ss}}{k_{Sm}}.
\end{equation}
Substituting the above expression for $c(R)$ into Eq. (\ref{J_ss1}), we arrive at
\begin{equation}
    J_{ss}=\left( k_{HBP}+\frac{k_{s}k_{b}}{k_{d}+k_{s}} \right) \left( c_{\infty}-\frac{J_{ss}}{k_{Sm}} \right).
\end{equation}
Eventually, we determine the steady-state flux by solving this equation. This leads to
\begin{equation}
    J_{ss}=\frac{k_{Sm} \left( k_{HBP}+\frac{k_{s}k_{b}}{k_{d}+k_{s}} \right)}{k_{Sm}+ k_{HBP}+\frac{k_{s}k_{b}}{k_{d}+k_{s}}} c_{\infty}=J_{Sm} \frac{k_{HBP}+\frac{k_{s}k_{b}}{k_{d}+k_{s}}}{k_{Sm}+ k_{HBP}+\frac{k_{s}k_{b}}{k_{d}+k_{s}}}.
\end{equation}

Now we take advantage of the relations in Eq. (\ref{J_ss}) to find the effective rate constant and trapping probability:
\begin{equation}\label{k1}
    k=\frac{J_{ss}}{c_{\infty}}=\frac{k_{Sm} \left( k_{HBP}+\frac{k_{s}k_{b}}{k_{d}+k_{s}} \right)}{k_{Sm}+ k_{HBP}+\frac{k_{s}k_{b}}{k_{d}+k_{s}}},
\end{equation}
and
\begin{equation}\label{Ptr1}
    P_{tr}=\frac{J_{ss}}{J_{Sm}}=\frac{k}{k_{Sm}}=\frac{k_{HBP}+\frac{k_{s}k_{b}}{k_{d}+k_{s}}}{k_{Sm}+ k_{HBP}+\frac{k_{s}k_{b}}{k_{d}+k_{s}}}.
\end{equation}
Finally, we find the effective trapping rate from Eq. (\ref{kappa_eff})
\begin{equation}\label{kappa_eff1}
    \kappa_{eff}=\kappa_{eff}^{HBP}+ \frac{1}{4 \pi R^{2}} \frac{k_{s}k_{b}}{k_{d}+k_{s}}.
\end{equation}

A particle diffusing on the surface of the sphere either dissociates and becomes unbound or is trapped by the disk. The probability of the second outcome, denoted by $P_{tr}^{(s)}$, is given by
\begin{equation}\label{Ptrs}
    P_{tr}^{(s)}=\frac{k_{s}}{k_{s}+k_{d}}.
\end{equation}
Introducing the notation $\kappa_{s}$ for the binding rate of unbound particles to the surface of the sphere outside the absorbing disk,
\begin{equation}
    \kappa_{s}=\frac{k_{b}}{4 \pi R^{2}},
\end{equation}
we can rewrite the effective trapping rate, Eq. (\ref{kappa_eff1}), as
\begin{equation}\label{kappa_eff2}
     \kappa_{eff}=\kappa_{eff}^{HBP}+ \kappa_{s} P_{tr}^{(s)}.
\end{equation}
We use the trapping probability $P_{tr}^{(s)}$ to write Eqs. (\ref{k1}) and (\ref{Ptr1}) in the form convenient for further analysis:
\begin{equation}\label{k2}
    k=\frac{k_{Sm}\left(k_{HBP}+k_{b}P_{tr}^{(s)}\right)}{k_{Sm}+k_{HBP}+ k_{b}P_{tr}^{(s)}},
\end{equation}
and
\begin{equation}\label{Ptr2}
    P_{tr}=\frac{k_{HBP}+k_{b}P_{tr}^{(s)}}{k_{Sm}+k_{HBP}+ k_{b}P_{tr}^{(s)}}.
\end{equation}

To summarize, main results of our analysis are the expressions for the effective rate constant, Eq. (\ref{k2}), the trapping probability, Eq. (\ref{Ptr2}), and the trapping rate, Eq. (\ref{kappa_eff2}).

The effective rate constant $k$ is a function of the geometric parameters $a$ and $R$ as well as the parameters that characterize the dynamics of the system: $D$, $D_{s}$, $k_{b}$ (or $\kappa_{s}$), and $k_{d}$. From Eqs. (\ref{k2}) and (\ref{Ptr2}), one can see that the effect of particle binding to the surface on the effective rate constant and trapping probability (at fixed values of $a$, $R$ and $D$) is determined by the term $k_{b}P_{tr}^{(s)}=k_{b}k_{s}/(k_{d}+k_{s})$. The rate constant $k$ is a monotonically increasing function of this parameter. As $k_{b}P_{tr}^{(s)}$ increase from zero (no effect of binding) to infinity (very strong binding), the effective rate constant grows from its minimum value $k_{min}=k_{Sm}k_{HBP}/(k_{Sm}+k_{HBP}) \simeq k_{HBP}$ to its maximum value $k_{max}=k_{Sm}$. Correspondingly, the effective trapping probability increases from $k_{HBP}/k_{Sm}$ to unity.

Note that the effective rate constant in Eq. (\ref{k2}) approaches zero, as $a \rightarrow 0$, much more slowly than $k_{HBP}$ given in Eq. (\ref{kHBP}). This is a consequence of the fact that the surface diffusion of the particles to the disk plays a dominant role in this limiting case. According to Eq. (\ref{k2}), we have
\begin{equation}
    k|_{a \rightarrow 0} \simeq \frac{k_{b}k_{s}}{k_{d}}=\frac{k_{b} D_{s}}{k_{d} R^{2} \left[2 \ln(2R/a)-1 \right]},
\end{equation}
whereas $k_{HBP}=4Da$ is proportional to the disk radius.

\section*{Numerical Tests of the Theory}

To test our analytical theory for a wide range of parameters, we employ Brownian Dynamics (BD) simulations. More specifically, our goal is to compute the mean lifetime of a Brownian particle diffusing in a spherical layer between two concentric spheres. The outer sphere of radius $R_{out}$ is a reflecting boundary for the particle. The inner sphere of radius $R$ ($R< R_{out}$) contains a small absorbing disk of radius $a$ ($a \ll R$). The particle can reversibly bind to the surface of the inner sphere outside the disk, and we assume that it diffuses on the surface with the same diffusivity $D$ as in the bulk. Thus the particle can reach the disk and be trapped either coming to the disk from the bulk or by surface diffusion. At $t=0$, the particle is not bound to the surface and its starting point is uniformly distributed over the surface of the inner sphere, including the disk area.


\begin{figure}
    \centering
    \includegraphics[clip,width=1.00\linewidth]{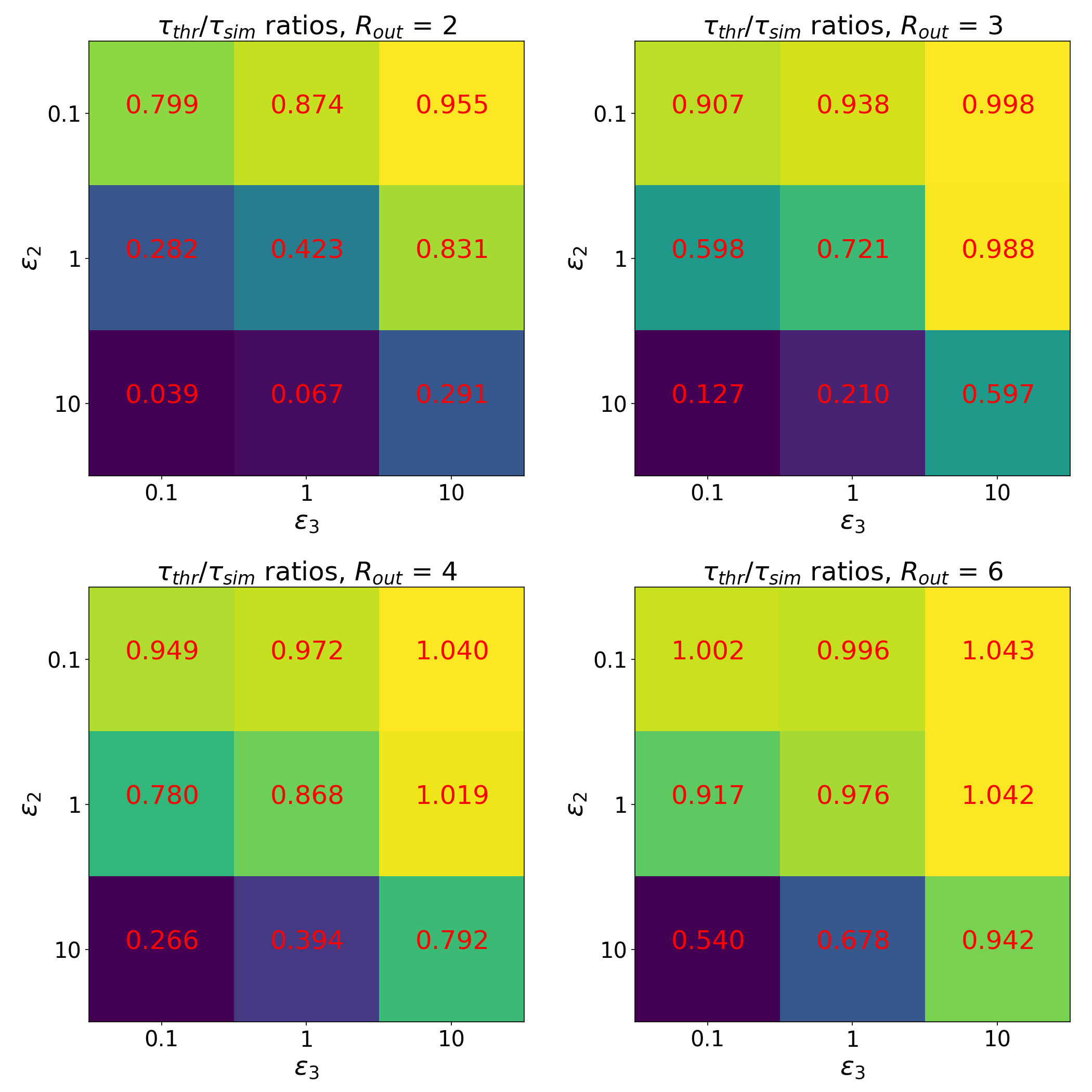}
    \caption{Comparison between numerical simulations and theoretical predictions for mean lifetimes of Brownian particles for $R_{out}=2, 3, 4$ and $6$.}
    \label{fig:times}
\end{figure}

The system is characterized by 6 dimensional parameters such as $a$, $R$, $R_{out}$, $D$, $k_{d}$, and $\kappa_{s}$. It seems convenient then to use $R$ as the unit of length and $R^{2}/D$ as the unit of time. So in our computer simulations we choose $R=D=1$. The mean particle lifetime $\tau_{sim}$ obtained from the computer simulations can be compared with its counterpart predicted by our analytical theory, $\tau_{theory}$. Then the measure of the success of our theoretical analysis is how close are these two times, i.e., we need to calculate the ratio $\tau_{theory}/\tau_{sim}$. 

The main idea of our approach is to replace the non-uniform surface of the inner sphere by homogeneous partially absorbing surface with the effective trapping rate $\kappa_{eff}$. The mean lifetime of a particle diffusing between partially absorbing and fully reflecting surfaces of radii $R$ and $R_{out}$, respectively, that starts from the partially absorbing surface, is given by\cite{dagdug2016boundary}
\begin{equation}
    \tau_{theory}=\frac{1}{3 \kappa_{eff} R} (R_{out}^{3} -R^{3}).
\end{equation}
This time can be rewritten in the dimensionless form as
\begin{equation}
    \widetilde{\tau}_{theory}=\tau \frac{D}{R^{2}}=\frac{1}{3 \widetilde{\kappa}_{eff}} \left[ \left(\frac{R_{out}}{R} \right)^{3}-1 \right],
\end{equation}
where $\widetilde{\kappa}_{eff}$ is the dimensionless effective trapping rate of the surface given by
\begin{equation}
    {\widetilde{\kappa}}_{eff}=\kappa_{eff} \frac{R}{D}.
\end{equation}

For numerical tests, it is convenient to present the dimensionless effective trapping rate as
\begin{equation}
    \widetilde{\kappa}_{eff}=\frac{a}{\pi R-a} +\frac{\varepsilon_{2}}{1+\varepsilon_{3}},
\end{equation}
where new dimensionless parameters are given by
\begin{equation}\label{epsilons}
    \varepsilon_{2}= \frac{k_{b}}{k_{Sm}}=\frac{\kappa_{s}R}{D}, \quad \varepsilon_{3}= \frac{k_{d}}{k_{s}}=k_{d} \frac{R^{2}}{D} \left[2 \ln(2R/a)-1 \right].
\end{equation}
Our idea is to test analytical predictions in computer simulations by varying $\varepsilon_{2}$ and $\varepsilon_{3}$ since they contain all relevant parameters of the process.

The comparison between theoretical predictions and the results of computer simulations for the surface-assisted binding process are presented in Fig. 2. One can see that pushing the outer boundary away from the surface (increasing $R_{out}$) significantly improves the agreement between the theory and simulations. This is an expected result because the theoretical arguments for substituting the originally heterogeneous surface with an effective homogeneous one are applicable for $R_{out}/R \gg 1$. One should also notice that theory works quite well for all ranges of parameters already for small $\varepsilon_{2}$ and large $\varepsilon_{3}$ that corresponds to weak bindings to the surface and frequent dissociations from the surface. In this case, the surface plays a relatively small role in the overall dynamics. The opposite limit of large $\varepsilon_{2}$ and small $\varepsilon_{3}$ presents the situation where the presence of the surface is crucial. In this case, the agreement between theory and simulations is not as good, although increasing $R_{out}$ clearly shows the right tendency.

\begin{figure}
   \centering
   \includegraphics[clip,width=0.90\linewidth]{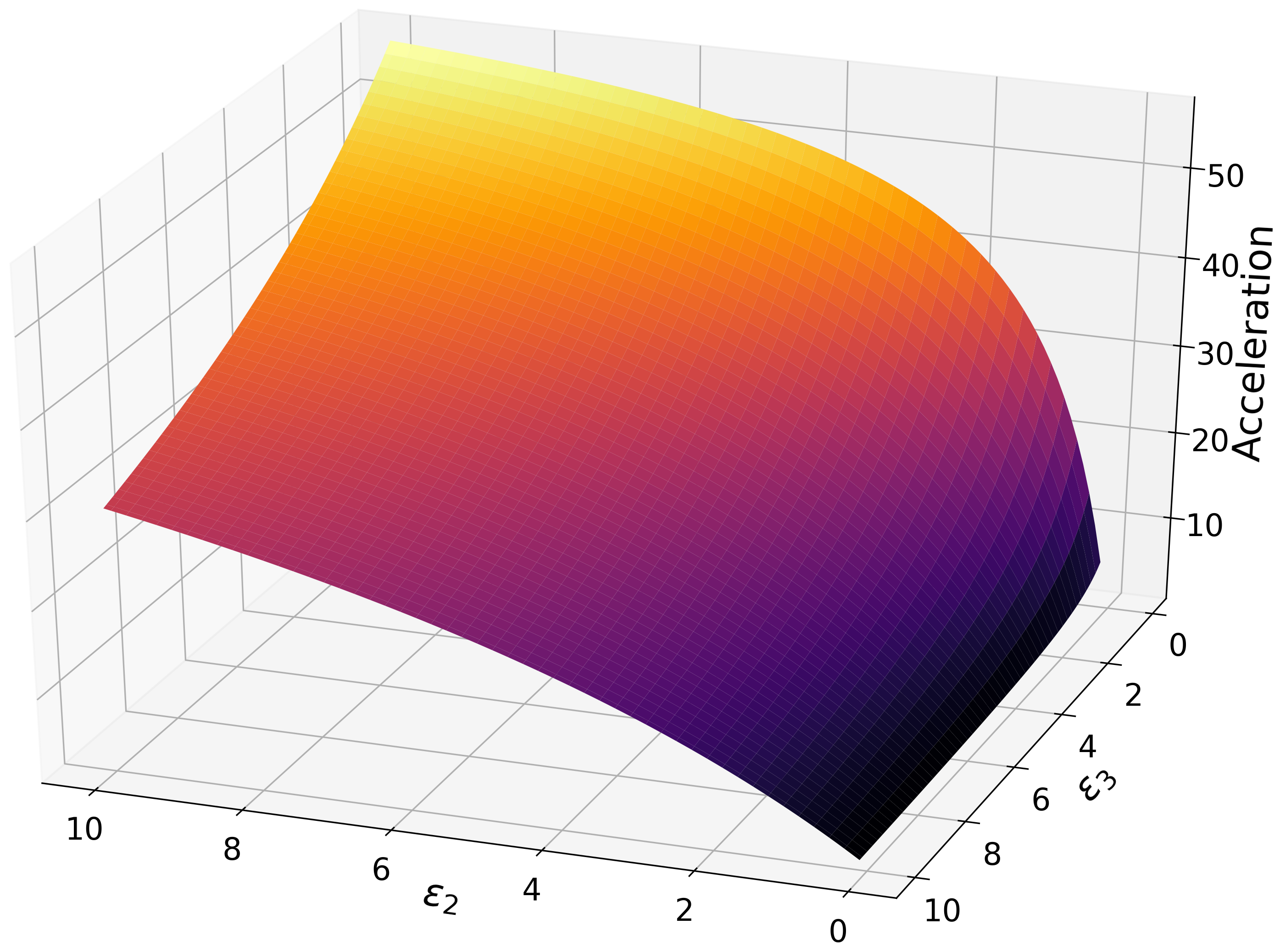}
    \caption{Acceleration of binding due to presence of nonspecific interactions for $R=1$, $a=0.05$ as a function of dimensionless parameters $\epsilon_2$ and $\epsilon_3$. Maximum theoretical acceleration for this system is $\gamma \sim 62.8$.}
   \label{fig:3d}
\end{figure}

To emphasize the important role of the surface in our system, we also consider an effective acceleration parameter $\gamma$ which is defined as a ratio of the effective trapping rate constant $k$, Eq. (31), to the Hill-Berg-Purcell rate constant $k_{HBP}$, Eq. (9), that corresponds to the fully inert sphere.  This ratio provides a quantitative measure of how the presence of the surface influences the overall trapping rate by the specific site. It can be shown that
\begin{equation}
  \gamma = \frac{k}{k_{HBP}}=\frac{1 + \frac{\pi R}{a}\frac{\epsilon_2}{1+\epsilon_3}}{1+\frac{a}{\pi R} + \frac{\epsilon_2}{1+\epsilon_3}}.
   \label{eq:gamma}
\end{equation}
Fig. \ref{fig:3d} presents this acceleration as a function of two dimensionless parameters $\epsilon_2$ and $\epsilon_3$ for the system with $a=0.05$. As given in Eqs. (\ref{epsilons}),  the parameter $\epsilon_2$ is proportional to the absorption constant ($\kappa_{s}$) and the parameter $\epsilon_3$ is proportional to the desorption rate constant ($k_d$). From the Fig. \ref{fig:3d}, one can see that the highest values of the acceleration are achieved for high values of $\epsilon_2$ and low values of $\epsilon_3$. The maximum theoretically possible acceleration can be simply estimated as
\begin{equation}
    \gamma_{max} \simeq \frac{k_{Sm}}{k_H}=\frac{\pi R}{a}.
    \label{eq:acc}
\end{equation}
This means that the surface effect is most pronounced when the active site is very small and the surface binding is strong.

\section*{Summary and Conclusions}

We developed a new theoretical framework to describe trapping of particles by an active site located on the surface. It generalizes a Smoluchovskii-Collins-Kimball approach by taking into account nonspecific interactions between the particles and the surface where the active site is located.  Our main idea is to replace the inhomogeneous surface by an effective homogeneous one. This allowed us to obtain a full description of the processes in the system. Importantly, we present an explicit analytical expression for the effective trapping rate. Our approximate theoretical method gives correct predictions in the limiting cases of strong binding and no binding to the surface. It was also tested using extensive BD computer simulations. The analysis suggests  that nonspecific interactions can significantly accelerate the trapping process in the systems with small active sites. 

\section*{Acknowledgments}

MM and ABK acknowledge the support from the Welch Foundation (C-1559), from the NSF (CHE-1953453 and MCB-1941106), and from the Center for Theoretical Biological Physics sponsored by the NSF (PHY-2019745). AMB and SMB were supported by the Intramural Research Program of the National Institutes of Health, {\it Eunice Kennedy Shriver} National Institute of Child Health and Human Development, and the Center for Information Technology.

\section*{Data Availability Statement}
The data that support the findings of this study are available from the corresponding author upon reasonable request.


\bibliography{biblio.bib}

\end{document}